\documentclass[aps,prl,twocolumn]{revtex4}
\usepackage{amsmath,bm,epsfig}

\def\Fbox#1{\vskip1ex\hbox to 8.5cm{\hfil\fboxsep0.3cm\fbox{%
  \parbox{8.0cm}{#1}}\hfil}\vskip1ex\noindent}  %%  {TEXT} in BOX

\newcommand{\B}[1]{{\bm{#1}}}%% Bold Roman & Greek Lower & Upper Case
\let \= \equiv \let\*\cdot \let\~\widetilde \let\-\overline

\begin{document}

\title{Universality of the Plastic Instability in Strained Amorphous Solids}
\author{Ratul Dasgupta$^1$, Smarajit Karmakar$^2$ and Itamar Procaccia$^1$}
\affiliation{$^1$Department of Chemical Physics, The Weizmann Institute of Science, Rehovot 76100, Israel, $^2$Dipartimento di Fisica,
Universita` di Roma ``La Sapienza", Piazzale Aldo Moro 2, 00185, Roma, Italy}

\date{\today}

\begin{abstract}
By comparing the response to external strains in metallic glasses and in Lenard-Jones glasses we find a quantitative universality of the
fundamental plastic instabilities in the athermal, quasistatic limit. Microscopically these two types of glasses are as different as one can imagine, the latter being determined by binary interactions, whereas the former by multiple interactions due to the effect of the electron gas that cannot be disregarded. In spite of this enormous difference the plastic instability is the same saddle-node bifurcation. As a result the statistics of stress and energy drops in the elasto-plastic steady state are universal, sharing the same system-size exponents.
\end{abstract}

\maketitle

The nature of the fundamental plastic instability in simple models of amorphous solids with binary interactions was elucidated
in recent years \cite{Lacks1999_JCP, Barrat2002_PRB, Lemaitre2004_PRL1, Lemaitre2006_PRE, Procaccia2009_PRE, Procaccia2010_PRE, Procaccia2011_PRE,Procaccia2010_PRL}. The fundamental plastic instability is best studied in athermal, quasi-static deformation of an amorphous solid such as to eliminate the effect of thermal fluctuations and finite strain rates. Contrary to some pictures suggesting that plastic instabilities occur at pre-determined `weak' sites called `shear transformation zones' \cite{79Arg,98FL}, it was discovered that the plastic instability
occurs as the vanishing of an eigenvalue of the Hessian matrix $\B H$ where the latter is defined as
\begin{equation}
H_{ij} \equiv \frac{\partial^2 U(\B r_1,\B r_2,\dots,\B r_N)}{\partial \B r_i \partial \B r_j} \ , \label{Hes}
\end{equation}
where $U$ is the total potential energy which is a function of the positions of all the $N$ particles in the system. Besides the three
obvious zero eigenvalues associated with Goldstone modes, all the other eigenvalues of $\B H$ are real positive as one expects from a real positive-definite symmetric matrix. The plastic instabilities exhibit a universal nature for all the studied binary glasses: at zero external strain ($\gamma=0$) all the eigenvalues of $\B H$ besides the Goldstones are positive, and the low lying eigenvalues are associated
with extended eigenfunctions. As the external strain is increased one eigenvalue begins to go down towards zero, and at the same time
its associated eigenfunction becomes localized. The eigenvalue, denoted $\lambda_P$, hits zero at $\gamma_P$ via a saddle node
bifurcation (such that a minimum in the potential energy landscape hits a saddle) and the way that it does so is universal, i.e.
\begin{equation}
\lambda_P \propto \sqrt{\gamma_P - \gamma} \ . \label{lamP}
\end{equation}
This scaling law has many interesting consequences. One immediate consequence is that the barrier $\Delta E$ between the minimum in which
the system resides and the saddle that it eventually collides with tends to zero like \cite{Procaccia2010_PRE1}
\begin{equation}
\Delta E\propto (\gamma_P-\gamma)^{3/2} \ . \label{DelE}
\end{equation}
Less obvious (and maybe more interesting) is the consequence of these scaling laws on the system size dependence of the statistics of
stress and energy drops in the elasto-plastic steady state. Once the stress level reaches the `yielding transition' and the system settles
in a steady state, every plastic instability is associated with a stationary statistics for the drops $\Delta \sigma$ in the stress and $\Delta U$ in the potential energy.
The long strain averages of these quantities scale with the system size according to  \cite{Lemaitre2004_PRL1, Lemaitre2006_PRE, Procaccia2009_PRE}.
\begin{eqnarray}
\langle\Delta U\rangle \sim N^{\alpha} ,\quad \langle\Delta \sigma\rangle \sim N^{\beta} \ . \label{scales}
\end{eqnarray}
It was shown that the numerical values of the exponents are determined by two ingredients. The first is the scaling laws (\ref{lamP}) and (\ref{DelE}). The second one is the assertion that after the yielding transition the probability to see a zero value of $\Delta E$ is not zero, in distinction from the unstrained solid in which this probability is zero, \cite{Procaccia2010_PRE2,09RS}. The two ingredients combined result in $\alpha=1/3$ and $\beta=-2/3$ universally for all binary glasses and in both 2 and 3 dimensions \cite{Procaccia2010_PRE2}. In addition, it was shown that the statistics of plastic events continue to be dominated by the fundamental plastic instabilities discussed here also for
finite strain rates and temperatures up to about 2/3 of $T_g$, see Ref. \cite{10CCL}.

In this Letter we study the fundamental plastic instability in a totally different class of glasses, i.e. metallic glasses where
the microscopic interaction is very different from binary since the conducting electron cloud mediates interactions beyond the
pair-wise. We study the response of a model of the metallic glass Cu$_{46}$Zr$_{54}$ in 2 and 3 dimensions using the Embedded Atom Model, and show quantitative universality with the binary glasses. In spite of the tremendous difference in microscopic interactions, the scaling laws (\ref{scales}) will be shown to repeat verbatim, with the very same scaling exponents.
\begin{figure}
\centering
\includegraphics[scale=0.25,angle=-90]{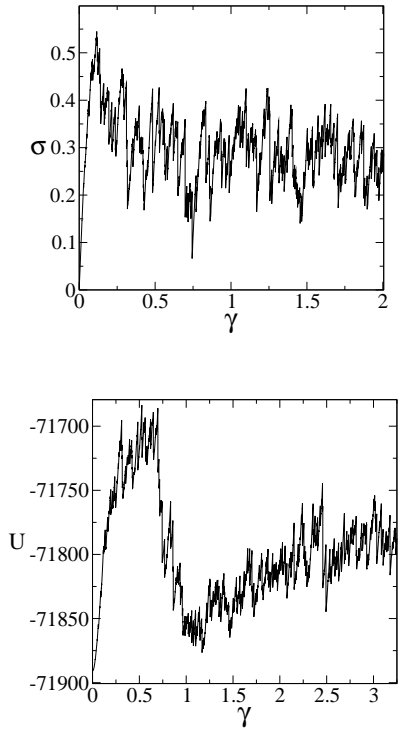}
\vskip 0.6 cm
\hskip -1.0 cm
\includegraphics[scale=0.25]{MetallicFig.1b.eps}
\caption{Upper panel: typical stress vs strain curve for a system of 4000 particles. Lower panel: a typical
potential energy vs strain curve for the same system as in the upper panel.}
\label{stressenergy}
\end{figure}

In order to simulate the consequences of the electron cloud we use a many-body potential. This potential represents the cohesive energy of an atom by the local electron density at that atom, the latter quantity itself being determined by the neighbors of the atom. A pure pair-wise interaction is added in order to represent the electrostatic repulsion (see \cite{Liu2008_PhyRep} and references therein for a more detailed description).  The expression for the total potential energy is
\begin{eqnarray}\label{U_tot1}
U &=& \sum_m\left(\sum_{n \neq m}\phi\left(r_{mn}\right)\right) + \sum_m F\left(\rho _m\right) \ , \\
\rho_m &=& \sum_{n \neq m} \psi\left(r_{mn}\right) \quad r_{mn}\equiv |\B r_n -\B r_m| \ .
\end{eqnarray}
Here $\phi$ is the pair-wise potential, $F\left(\rho _{m}\right)$ is the many-body term representing the energy of the $m$th atom and $\rho _m$ is a measure of the local electron density at the location of the $m$th atom, due to contributions from the neighboring atoms \cite{Liu2008_PhyRep}. The functional form of $F\left(\rho _{m}\right)$ is taken as customary in the literature, cf. \cite{Legrand1989_PhilMag, Rosato1993_PRB} i.e. the semi-empirical approximation $F\left(\rho _{m}\right) \sim -\sqrt{\rho _m}$. We employed the functional forms of $\phi$ and $\psi$ from Ref. \cite{Goddard2005_PRB} where fits were obtained for this model to density functional quantum mechanical calculations, for the case of the bulk metallic glass Cu$_{46}$Zr$_{54}$. For the sake
of computational speed we cut-off the functions $\Phi$ and $\Psi$ to go to zero smoothly (up to second derivative) at a finite distance.
Explicitly, we used the following functional forms \cite{Goddard2005_PRB},
\begin{eqnarray}\label{phi_psi}
\phi &=& \frac{1}{2} \epsilon_{mn}\Big[e^{-p_{mn}\left(\frac{r_{mn}}{\sigma_{mn}} -1\right)} + A_1^{mn} \nonumber\\
&+& B_1^{mn}\left(\frac{r_{mn}}{\sigma_{mn}}\Big) + C_1^{mn}\left(\frac{r_{mn}}{\sigma_{mn}}\right)^2\right] \ , \label{phi}\\
\psi &=& \left(c^{mn}\right)^2\Big[ e^{-2q^{mn} \left(\frac{r_{mn}}{\sigma_{mn}} -1 \right)}  + A_2^{mn} \nonumber\\&+& B_2^{mn}\left(\frac{r_{mn}}{\sigma_{mn}}\right) + C_2^{mn}\left(\frac{r_{mn}}{\sigma_{mn}}\right)^2\Big] \ . \label{psi}
\end{eqnarray}
Here the coefficients $A$, $B$ and $C$ have been added to achieve smooth first and second derivatives of $\phi$ and $\psi$ at the cut-off.  The values of the parameters $p$, $q$ $\epsilon$, $\sigma$ and $c$ in equations (\ref{phi}) and (\ref{psi}) are taken from \cite{Goddard2005_PRB}. Note that cutting off the binary functions at a finite distance does not remove the multi-body interaction because the force between
any pair or particles is configuration dependent through the second term in the total energy. For our purposes it suffices to smooth
the potential up to second derivative since we are interested in the Hessian matrix.  We use a (scaled) value of cut-off $1.707$ corresponding to an absolute value of $4.5 \AA$ used in \cite{Goddard2005_PRB}. The scaled value of density is $0.951$ obtained from the data in \cite{Goddard2005_PRB} corresponding to $2000$ atoms occupying a volume of $38.5$ $nm^3$ (close to $T_g$ in \cite{Goddard2005_PRB}). All the simulations reported here are in three dimensions, keeping a constant density in a 3-dimensional box with periodic boundary conditions in all directions.
\begin{figure}
\includegraphics[scale=0.35, angle = -90]{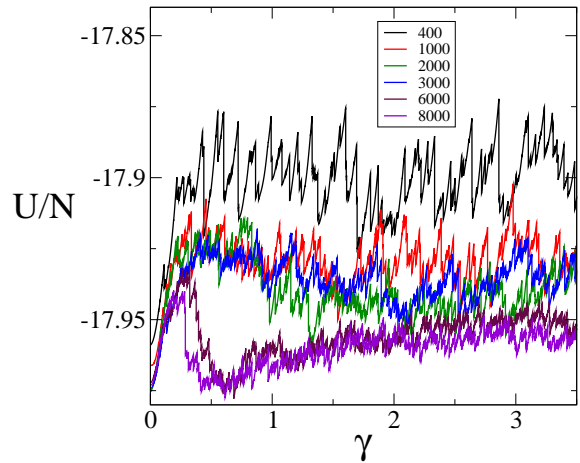}
\caption{ Color online: Potential Energy per particle versus strain for systems of different sizes between N$=400$, and $N=8000$. Note the energy peak and
then dip before the elasto-plastic steady state is established, this seems particular to the metallic glass and not commonly seen in binary glasses.}
\label{envsN}
\end{figure}

We have put our model metallic glass under a shear strain using the usual algorithm for an athermal quasi-static process \cite{Procaccia2009_PRE}.
Fig. \ref{stressenergy} demonstrates a typical stress vs. strain and potential energy vs. strain curves for a metallic glass
consisting of 4000 particles.
We see the usual linear regime in the upper panel where the stress is linear in the strain, and the corresponding regime
where the potential energy is quadratic in the strain in the lower panel. Every drop in the curves represents a plastic
failure, and we are interested in the statistics and system size dependence of those. Indeed, in Fig. \ref{envsN} we show the potential
energy {\em per particle} vs strain for systems of different sizes from $N=400$ to $N=8000$. We see the decline in the amplitude of the
fluctuations but also a small reduction in the average energy in the elasto-plastic steady state. Note the energy peak and
then dip before the elasto-plastic steady state is established, this seems particular to the metallic glass and not commonly seen in binary glasses.

\begin{figure}
\includegraphics[scale=0.24]{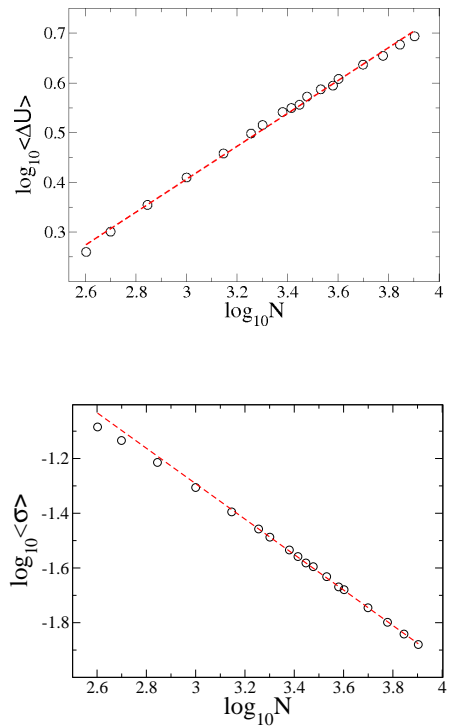}
\vskip 1.0 cm
\includegraphics[scale=0.24]{MetallicFig.3b.eps}
\caption{Color online: The scaling of $\langle \Delta U\rangle $ (upper panel) and $\langle\Delta\sigma \rangle$ (lower panel) with the system size $N$. Symbols indicate numerical data while the dashed line is a fit. The dashed curves are power laws with exponents $\alpha=0.330$ and
$\beta=-0.647$ respectively, in good agreement with the power laws (\ref{scales}) which predict universal exponents $\alpha=1/3$ and $\beta=-2/3$.}
\label{scalelaw}
\end{figure}

The scaling laws with respect to the system size are examined in Figs. \ref{scalelaw}.
\begin{figure}
\includegraphics[scale=0.25]{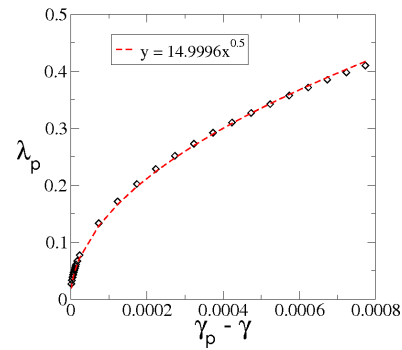}
\caption{Color online: The lowest eigenvalue of the Hessian matrix as it drops to zero at a typical plastic event in the elasto-plastic steady state.
The agreement with the saddle-node bifurcation scenario (dashed red curve) seems perfect.}
\label{eig}
\end{figure}
\begin{figure}
\includegraphics[scale=0.35]{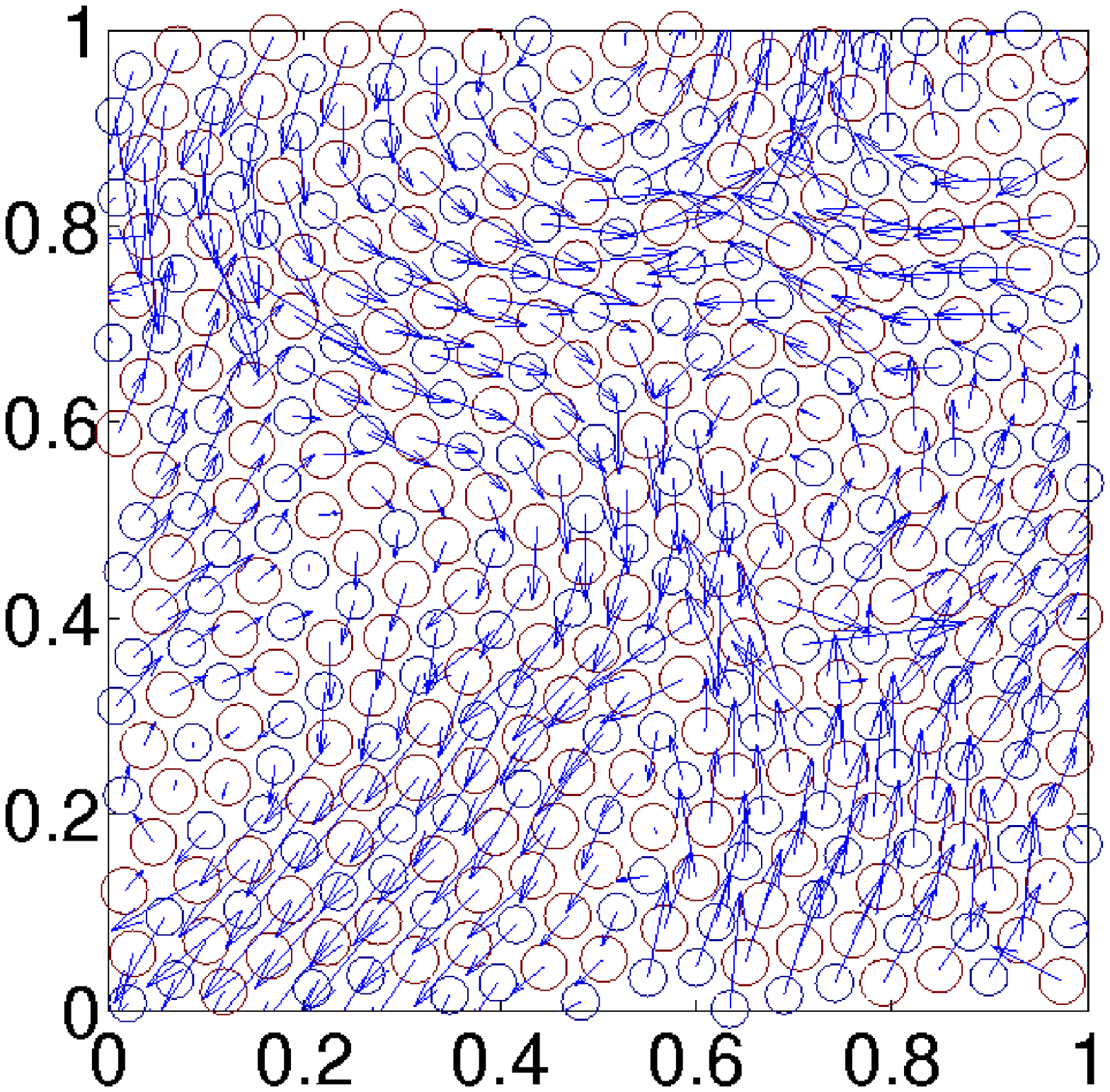}
\includegraphics[scale=0.35]{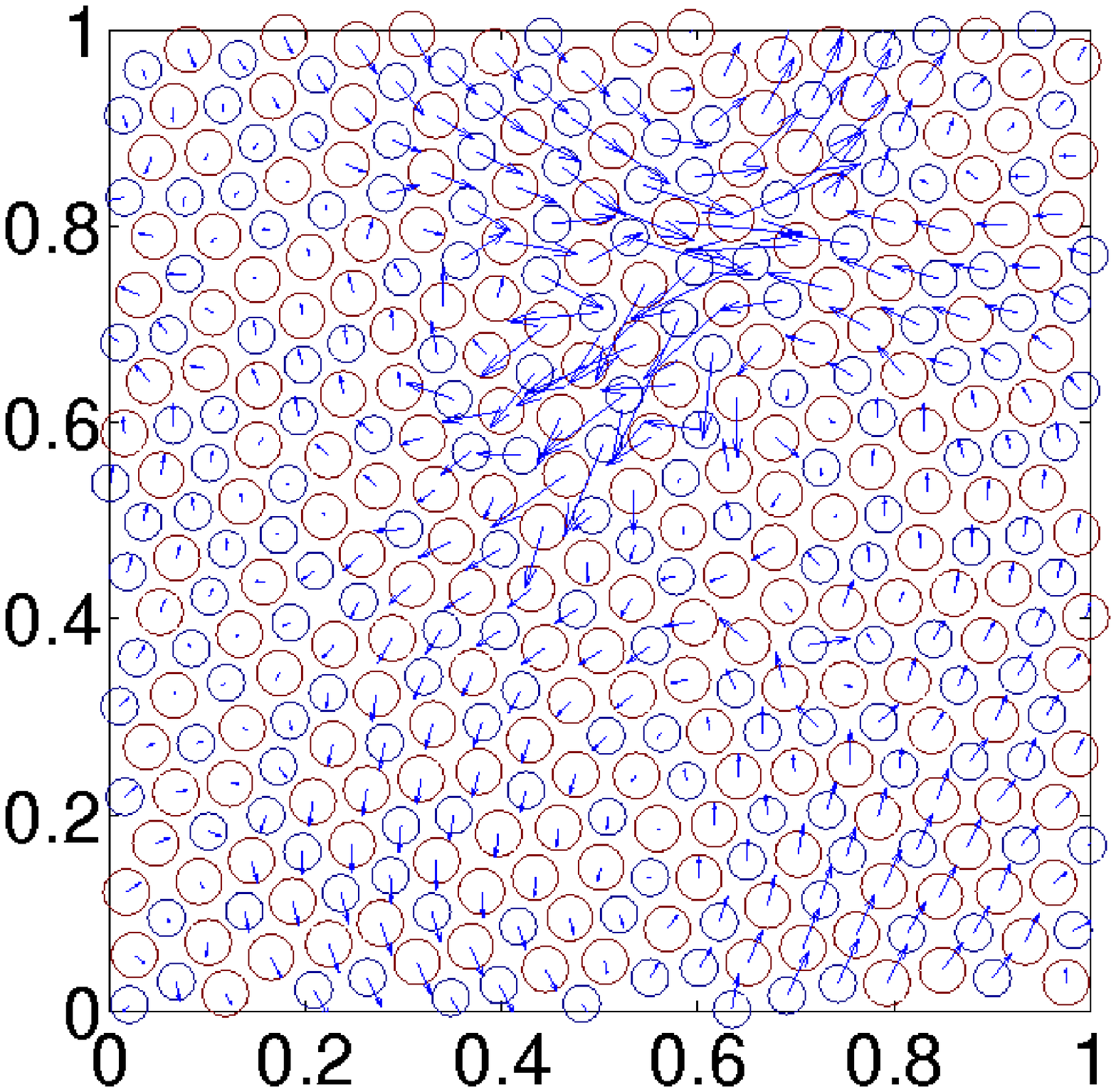}
\caption{Color online: A 2-dimensional demonstration of the nature of the plastic instability. Upper panel: at $\gamma=0$ the eigenfunction associated with the smallest non-zero eigenvalue is perfectly extended without any sign of the position of the incipient instability. Lower Panel:
the localized eigenfunction associated with the very same eigenvalue when it approaches zero. (see Fig. \ref{eig}).}
\label{eigfun}
\end{figure}
We see that Eqs. \ref{scales} are reproduced by the data with the scaling exponents being very close to the theoretical values
$\alpha=1/3$ and $\beta=-2/3$. This demonstration of universality is determined by the fundamental plastic instability which, in
this case of metallic glass, in spite of the huge difference in microscopic interactions, is the same saddle node bifurcation as
in all the simple cases of binary glasses. To show that this is so we need to diagonalize the Hessian whose computation is
somewhat delicate in the present case. The analytic form of the Hessian for our Embedded Atom Model is a bit cumbersome to write
here explicitly, but it can be found in \cite{Hessian}. A typical calculation of the lowest eigenvalue for the case $N=400$ is shown in
Fig. \ref{eig}. We see how the eigenvalue goes to zero at a value $\gamma_P$ precisely as expected for a saddle node bifurcation, i.e.
according to Eq. (\ref{lamP}).

The conclusion of this study is two-fold. First, we learn that the nature of the precise microscopic model does not matter much
in determining the fundamental plastic instability of amorphous solids. Once coarse-grained, or up-scaled to provide us with an
energy landscape, the plastic instability occurs when the strain distorts the landscape such that a local minimum collides with
a near-by saddle. Such a collision is a generic saddle node bifurcation that will be seen as an eigenvalue of the Hessian matrix
nearing zero with a square-root singularity. The power of genericity of bifurcations exceeds the details of the inter-particle
potential. Secondly, we learn once more that it is futile to seek `shear transformation zones' in the material. At rest, unstrained, the material does not exhibit any pre-existing local weak points where the plastic event will take place. The plastic event is a result of the straining process, in which an extended eigenfunction of the Hessian matrix gets localized at the position of the plastic event.
This lesson is as true for simple glasses as for the metallic glass. We conclude this letter by showing the eigenfunction associated
with the lowest eigenvalue at $\gamma=0$ and at $\gamma_P$ in a 2-dimensional metallic glass (for ease of display,  cf. Fig. \ref{eigfun}). In equilibrium the eigenfunction is perfectly extended, and there is nothing in it that can predict the position of the incipient instability. Once localized, the eigenfunction exhibits the typical
quadrupolar structure that is associated with the well-known Eshelby inclusion \cite{57Esh}, which is universal for elastic materials
independent of the microscopic details.


\begin{thebibliography}{99}
\bibitem{Lacks1999_JCP}
%Eigenfrequency going to zero paper
D. L. Malandro and D. J. Lacks, J. Chem. Phys. {\bf 110}, 9 (1999).
%%
\bibitem{Barrat2002_PRB}
A. Tanguy, J.P. Wittmer, F. Leonforte and J-L Barrat , Phys. Rev. B {\bf 66}, 174205 (2002).
%%
\bibitem{Lemaitre2004_PRL1}
C. E. Maloney and A. Lema\^{i}tre, Phy. Rev. Lett. {\bf 93}, 016001,195501 (2004).

%%
\bibitem{Lemaitre2006_PRE}
%Expanded version of PRL2
C. E. Maloney and A. Lema\^{i}tre, Phy. Rev. E {\bf 74}, 016118 (2006)

\bibitem{Procaccia2010_PRE}
%%Nonlinear Elastic constants
S. Karmakar, E. Lerner and I. Procaccia, Phy. Rev. E {\bf 82}, 026105 (2010)
%%
\bibitem{Procaccia2011_PRE}
%%Athermal solids exist ? paper
H. G. E Hentschel, S. Karmakar, E. Lerner and I. Procaccia, Phy. Rev. E {\bf 83}, 061101 (2011)

\bibitem{Procaccia2009_PRE}
%%Exponent paper
E. Lerner and I. Procaccia, Phy. Rev. E {\bf 79}, 066109 (2009)

\bibitem{Procaccia2010_PRL}
%%Predicting plastic flow events
S. Karmakar, A. Lema\^{i}tre, E. Lerner and I. Procaccia, Phy. Rev. Lett. {\bf 104}, 215502 (2010).
%%

\bibitem{79Arg}
A.S. Argon, Acta Metall. {\bf 27}, 47 (1979).

\bibitem{98FL}
M.L. Falk and J.S. Langer, Phys. Rev. E {\bf 57}, 7192 (1998).

\bibitem{Procaccia2010_PRE1}
%%Statistical physics of the elasto-plastic steady states
S. Karmakar, E. Lerner, I. Procaccia and J. Zylberg Phy. Rev. E {\bf 82}, 031301 (2010).
%%
\bibitem{Procaccia2010_PRE2}
%%Yielding transition paper
S. Karmakar, E. Lerner and I. Procaccia, Phy. Rev. E {\bf 82}, 055103(R) (2010).

\bibitem{09RS}
D. Rodney and C. Schuh, Phys. Rev. Lett, {\bf 102}, 235503 (2009).

\bibitem{10CCL}
J. Chattoraj, C. Caroli, and A. Lema\^{i}tre, Phys. Rev. Lett. {\bf 105}, 266001 (2010).
%%

\bibitem{Liu2008_PhyRep}
J. H Li {\sl et al.} , Phy. Rep. {\bf 455}, (2008).
%%

%%
\bibitem{Legrand1989_PhilMag}
V. Rosato, M. Guillope and B. Legrand, Philos. Mag. A, {\bf 59}, 321, (1989).
%%
\bibitem{Rosato1993_PRB}
F. Cleri and V. Rosato, Phys. Rev. B., {\bf 48}, 22, (1993).

\bibitem{Goddard2005_PRB}
G. Duan, D. Xu, Q. Zhang, G. Zhang, T. Cagin, W. L. Johnson, and W. A. Goddard, III, Phy. Rev. B, {\bf 71}, 224208, (2005), {\bf 74}, 019901, (2006).

\bibitem{Hessian}
http://www.weizmann.ac.il/chemphys/cfprocac/~(papers on line \#177).

\bibitem{57Esh}
J. D. Eshelby,   Proc. R. Soc. Lond. A , {\bf 241}, 376 (1957).
\end{thebibliography}
\end{document}